\documentclass[preprint,12pt]{elsarticle}

\usepackage{amssymb}
\usepackage{amsmath}
\usepackage{amsfonts}

\usepackage{multirow}
\usepackage{booktabs}

\usepackage{listings}

\usepackage[svgnames]{xcolor}

\usepackage{url}

\usepackage{csquotes}

\usepackage{algorithmic}

\usepackage{booktabs}
\usepackage{graphicx}

\usepackage[htt]{hyphenat}

\definecolor{diffstart}{named}{Grey}
\definecolor{diffincl}{named}{Green}
\definecolor{diffrem}{named}{OrangeRed}
\definecolor{codegreen}{rgb}{0,0.6,0}
\definecolor{codegray}{rgb}{0.5,0.5,0.5}
\definecolor{codepurple}{rgb}{0.58,0,0.82}
\definecolor{backcolour}{rgb}{0.95,0.95,0.92}

\lstdefinelanguage{Ini}
{
    basicstyle=\ttfamily\small,
    columns=fullflexible,
    morecomment=[s][\color{Orchid}\bfseries]{[}{]},
    morecomment=[l]{\#},
    morecomment=[l]{;},
    commentstyle=\color{codegray}\ttfamily,
    morekeywords={},
    otherkeywords={=,:},
    keywordstyle={\color{codegreen}\bfseries}
}

\lstdefinelanguage{Python}{
    basicstyle=\ttfamily\footnotesize,
    morecomment=[f][\color{diffstart}]{@@},
    morecomment=[f][\color{diffincl}]{+\ },
    morecomment=[f][\color{diffrem}]{-\ },
    keywords={def,if,for,in, self, print},
}

\lstdefinestyle{mystyle}{
    backgroundcolor=\color{backcolour},   
    commentstyle=\color{codegreen},
    keywordstyle=\color{magenta},
    numberstyle=\tiny\color{codegray},
    stringstyle=\color{codepurple},
    basicstyle=\ttfamily\scriptsize, %
    breakatwhitespace=false,         
    breaklines=true,                 %
    captionpos=b,                    
    keepspaces=true,                 
    numbers=left,                    
    numbersep=5pt,                  
    showspaces=false,                
    showstringspaces=false,
    showtabs=false,                  
    tabsize=2
}

\journal{Information of Software Technology}

\makeatletter
\def\theaffn{\ifcase\c@affn\or $\spadesuit$\or $\clubsuit$\or $\heartsuit$\or $\diamondsuit$\fi}

\makeatother
\renewcommand{\rq}[1]{\textbf{RQ#1}}

\begin{document}

\begin{frontmatter}

\title{Automating Detection and Root-Cause Analysis of Flaky Tests in Quantum Software}

\author[1]{Janakan~Sivaloganathan\fnref{equal}}
\ead{jsiva@torontomu.ca}
\author[2]{Ainaz~Jamshidi\fnref{equal}}
\ead{ainazj1@umbc.edu}
\author[1]{Andriy~Miranskyy}
\ead{avm@torontomu.ca}
\author[2]{Lei~Zhang}
\ead{leizhang@umbc.edu}

\fntext[equal]{Equal contributions.}

\affiliation[1]{organization={Department of Computer Science, Toronto Metropolitan University}, city={Toronto}, country={Canada}}
\affiliation[2]{organization={Department of Information Systems, University of Maryland, Baltimore County}, city={Baltimore County}, country={United States of America}}

\begin{abstract}

\textit{Context:} Like classical software, quantum software systems rely on automated testing. However, their inherently probabilistic outputs make them susceptible to quantum flakiness---tests that pass or fail inconsistently without code changes. Such quantum flaky tests can mask real defects and reduce developer productivity, yet systematic tooling for their detection and diagnosis remains limited.

\textit{Objective:} This paper presents an automated pipeline to detect flaky-test-related issues and pull requests in quantum software repositories and to support the identification of their root causes. We aim to expand an existing quantum flaky test dataset and evaluate the capability of Foundation Models (FMs), particularly Large Language Models (LLMs), for flakiness classification and root-cause identification.

\textit{Method:} Building on a prior manual analysis of 14 quantum software repositories, we automate the discovery of additional flaky test cases using LLMs and cosine similarity. We further evaluate a variety of LLMs from OpenAI GPT, Meta LLaMA, Google Gemini, and Anthropic Claude suites for classifying flakiness and identifying root causes from issue descriptions and code context. Classification performance is assessed using standard performance metrics, including F1-score.

\textit{Results:} Using our pipeline, we identify 25 previously unknown flaky tests, increasing the original dataset size by 54\%. The best-performing model, Google Gemini 2.5 Flash, achieves an F1-score of 0.9420 for flakiness detection and 0.9643 for root-cause identification, demonstrating that LLMs can provide practical support for triaging flaky reports and understanding their underlying causes in quantum software.

\textit{Conclusion:} Our results indicate that LLMs show promise for maintaining quantum software quality by automating the detection and diagnosis of quantum flaky tests. The expanded dataset and automated pipeline contribute reusable artifacts for the quantum software engineering community. Future work will focus on improving detection robustness and developing automated fixes for quantum flaky tests.

\end{abstract}


\end{frontmatter}

\section{Introduction}
\label{sec:introduction}

\begin{figure}[!htp]
  \begin{center}
  \includegraphics[width=0.6\columnwidth]{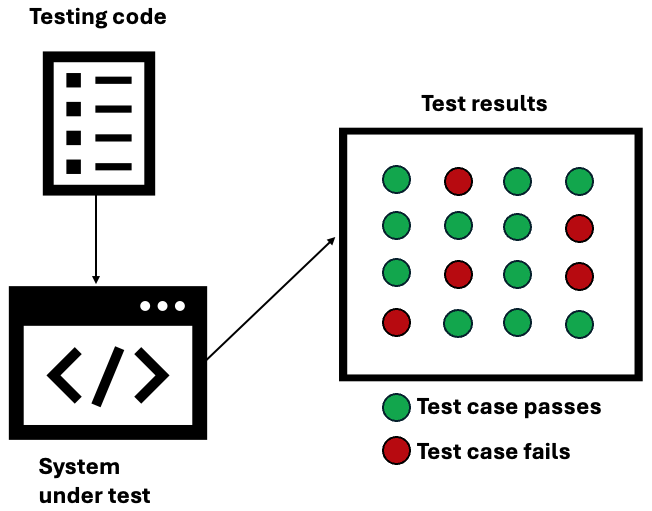}
  \end{center}
  \caption{A flaky test can generate random test results without any changes in code.}
  \label{fig:flaky}
\end{figure}

Flaky tests, which fail or pass inconsistently without any changes to the code under test (Fig.~\ref{fig:flaky}), pose significant challenges for software maintenance and reliability~\cite{luo2014empirical}. They can mislead developers by producing unreliable signals during regression testing. Flaky tests are often viewed as bugs in the test infrastructure: because they are difficult to reproduce, developers may postpone investigation or ignore intermittent failures, which can accumulate as technical debt and ultimately surface as production issues. Reproducing flakiness typically requires re-executing the same test many times, incurring substantial computational cost. At Google, for example, Luo et al.\ reported that in 2014, 73K out of 1.6M (4.56\%) test failures were attributed to flaky tests~\cite{luo2014empirical}; later reports also indicate that flakiness remained non-trivial at scale (e.g., 1.5\% of 4.2M tests in 2017)~\cite{micco2017state, memon2017taming}.

With the rapid growth of quantum computing technology and quantum software, the development, maintenance, and evolution of quantum software have attracted increasing attention and introduced new challenges to the software engineering community, contributing to the emergence of \emph{quantum software engineering}. In classical software engineering, flaky tests have been widely studied, including empirical investigations~\cite{luo2014empirical, gruber2021empirical}, flakiness detection techniques and tools~\cite{lam2019idflakies, silva2020shake, bell2018deflaker, alshammari2021flakeflagger, verdecchia2021know}, categorization approaches~\cite{akli2023flakycat}, and mitigation strategies~\cite{barbosa2022test, habchi2022qualitative}.

However, quantum flakiness remains underexplored. Our prior studies~\cite{zhang2023identifying,zhang2024automated,kaur2025identifying} have identified ``quantum flakiness'' in multiple open-source quantum projects. Similar to classical flaky tests, quantum flakiness manifests as inconsistent pass/fail outcomes during testing without code changes. However, quantum flakiness brings distinct challenges due to the probabilistic nature of quantum computation and the realities of quantum execution environments as follows. 
\begin{enumerate}
    \item \textbf{Different dominant root causes.} Empirical evidence suggests that randomness is a leading root cause of quantum flakiness~\cite{zhang2023identifying}, whereas classical flakiness is more commonly associated with concurrency and asynchronous behaviors~\cite{luo2014empirical,parry2021survey}.
    \item \textbf{Noise-driven variability.} In addition to algorithmic probabilism, quantum noise (e.g., decoherence and gate errors) can introduce further variability in outcomes, making it harder to distinguish genuine regressions from hardware-induced fluctuations.
    \item \textbf{High cost of reproduction on real hardware.} Diagnosing flakiness often requires repeated executions; on real quantum hardware, these re-runs can be substantially more expensive and constrained than on classical infrastructure (e.g., IBM quantum platform offers pay-per-use pricing at \$96 USD per minute with limited queue availability), amplifying the practical impact of flaky tests.
\end{enumerate} 

In our previous empirical study~\cite{zhang2023identifying}, we adopt keyword (e.g., flaky and flakiness) searches in issue reports (IRs) and pull requests (PRs) of 12 quantum software systems (such as IBM Qiskit~\cite{Qiskit2019} and NetKet~\cite{carleo2019netket}) and identify 46 unique quantum flaky tests.\footnote{We examined 14 quantum software repositories, but found flaky-test-related IRs and PRs in 12 of them.} We then identify and categorize eight types of flakiness and seven common fixes. Randomness is the most common cause of quantum flakiness, and the most common solution is to fix pseudo-random number generator (PRNG) seeds. %

Our findings in~\cite{zhang2023identifying} are constrained by the limitations inherent in the vocabulary-based method. In addition, we manually examine and identify all flaky tests, which is time-consuming. %
In this work, leveraging existing studies from classical software engineering, we explore a more effective and efficient flaky test detection technique for quantum software using IRs, PRs, and code contexts by answering the following research questions. 
\begin{itemize}
    \item \rq{1}: How can we detect new flaky tests based on the existing dataset?
    \item \rq{2}: In the new dataset, what are the common root causes and fixes for quantum flakiness?
    \item \rq{3}: How can we automatically detect if a given IR or PR is related to a flaky test?
    \item \rq{4}: How can we automatically detect if a given IR or PR is related to a flaky test with additional code context?
    \item \rq{5}: How can we automatically identify the root cause of a flaky test?
\end{itemize}

Our main \textbf{contributions} are as follows. We have 1) enriched the existing dataset of flaky tests~\cite{zhang2023identifying} by adding 54\% more observations, as well as including the buggy code causing flakiness and the corresponding fixes, which were absent from the original dataset (the extended dataset\footnote{The dataset may be leveraged by the community to build better models for flaky test detection and analysis.} is publicly available via~\cite{sivaloganathan_2026_18642687}); 2) developed a procedure to semi-automatically detect flaky test-related IRs and PRs by mining software repositories using embedding transformers and cosine similarity; 3) proposed an approach to detect flaky-test-related IRs and PRs using LLMs with zero, and few-shot prompting automatically.

As an initial step to deal with the challenges of quantum flakiness, we begin by analyzing both IRs/PRs and source code to detect flaky tests. Our future plans include but are not limited to dynamic testing (e.g., test re-execution) and extending the code analysis. By automating the process of flaky test detection and categorization, our ultimate goal is to improve the reliability and maintainability of quantum software systems.

This paper is organized as follows. Section~\ref{sec:related-work} reviews related work and identifies research gaps. Section~\ref{sec:data} describes dataset construction, extension, and root-cause labeling, answering RQs~1--2. Section~\ref{sec:automation} presents our automated pipeline for detecting quantum flaky tests and identifying their root causes using LLMs with textual and code context. Section~\ref{sec:results} reports the experimental results and answers RQs~3--5. Section~\ref{sec:threats} discusses threats to validity, and Section~\ref{sec:conclusions} concludes with directions for future work.

\section{Related Work}\label{sec:related-work}

\subsection{Classical Flaky Test Detection}

Flaky test detection has been an active and well-studied area of research in classical software engineering, with many techniques proposed to identify and mitigate flakiness.

The most straightforward techniques for automatically detecting flaky tests rely on repeatedly re-running tests under different conditions. iDFlakies~\cite{lam2019idflakies} executes test suites in randomized orders to detect order-dependent flaky tests. Shaker~\cite{silva2020shake} perturbs thread scheduling during test execution to expose concurrency-induced flakiness. These dynamic approaches can effectively reveal nondeterministic behaviors. However, they require multiple test executions, which can be time-consuming and resource-intensive, especially for large test suites and complex systems.

To reduce this overhead, researchers have proposed static and learning-based approaches for predicting flaky tests without repeated execution. FlakeFlagger~\cite{alshammari2021flakeflagger} uses static code metrics and change-history features with traditional machine learning models, such as $k$-Nearest Neighbors ($k$-NN), to predict test flakiness. Similarly, FLAST~\cite{verdecchia2021know} applies $k$-NN with historical test data to identify flaky tests and their potential causes, such as test smells and anti-patterns. While these ML-based methods reduce execution cost, they rely on supervised learning and require large, high-quality labeled datasets. Such data is scarce in quantum software projects. Moreover, the probabilistic nature of quantum programs further complicates feature design and model generalization.

With the emergence of LLMs, recent work has explored their use for flaky test detection in classical software. Flakify~\cite{fatima2022flakify} leverages LLMs to predict flaky tests by analyzing natural language descriptions and code changes. This approach reduces reliance on extensive labeled data. FlakyCat~\cite{akli2023flakycat} formulates flaky test categorization as a few-shot learning (FSL) problem and uses CodeBERT representations with Siamese networks for multi-class classification. These studies demonstrate that LLMs and FSL can capture semantic signals beyond handcrafted features.

\subsection{Quantum Flaky Test Detection}

The rapid growth of quantum computing has increased attention on quantum software development, maintenance, and evolution. This growth introduces new challenges for the software engineering community and motivates the emerging field of quantum software engineering. Quantum flakiness remained largely unexplored due to the lack of available datasets. 

In our initial study~\cite{zhang2023identifying}, we conducted the first systematic empirical analysis of flaky tests in quantum software across 14 open-source repositories. We identified 46 unique flaky test reports from 12 projects and categorized eight root causes and seven fix patterns. Randomness emerged as the dominant cause, in contrast to classical software, where concurrency and asynchronous waits are more prevalent. This study established the existence and practical relevance of quantum flakiness and produced the first curated dataset. However, detection relied on keyword-based search and manual triage, which limits recall and scalability.
Building on this empirical foundation, we outlined challenges and a research agenda for automating the detection of flaky-test-related bug reports in quantum software~\cite{zhang2024automated}. We framed flakiness detection as a supervised or semi-supervised learning task over issue report text and metadata, and highlighted challenges such as data scarcity, class imbalance, and noisy labels.

In our subsequent work~\cite{kaur2025identifying}, we proposed a feature-based machine learning pipeline to detect flaky tests in quantum code. This approach operates at the code level by extracting lexical features from Python files and training classical ML models, including XGBoost, Decision Trees, and Random Forests. We showed that tree-based models outperform other classifiers on both balanced and imbalanced datasets. We also extended the original quantum flakiness dataset with non-flaky examples to enable supervised learning. While this approach advances automated detection beyond keyword matching, it focuses on static code artifacts. It does not leverage natural language signals from issue reports and pull requests. It also does not address automated root cause identification from combined textual and code context.

This paper extends prior works by introducing an automated tool for flaky test detection and root cause analysis in quantum software. Unlike earlier approaches, we expand the existing quantum flakiness dataset using embedding-based similarity search to identify previously unknown flaky-test-related issues and pull requests. We employ LLMs to automatically (1) classify whether an IR or a PR is related to flaky tests, and (2) diagnose the root cause of the flakiness by jointly reasoning over the natural language description and the corresponding code context. This approach addresses key challenges in quantum software engineering, including the scarcity of labeled data and the probabilistic nature of quantum programs. This work represents a step toward practical, scalable automation for flaky test detection and diagnosis in quantum software engineering.

\section{Dataset}
\label{sec:data}

In this section, we answer \textbf{RQ1}, building on our previous manual analysis~\cite{zhang2023identifying}, which, to the best of our knowledge, provides the only dataset focused on quantum flaky tests. Our findings (designed to detect IRs and PRs related to flaky tests), though tailored to quantum software, also have the potential to detect reports of flaky tests in classical software.

\subsection{Data Expansion and Initial Analysis}

\begin{table*}[thb!]
    \centering
    \caption{Statistics of quantum software repositories with flaky tests. Three right-most columns are as follows: count of closed issue reports (Column $T$), count of closed flaky test reports (Column $F$), and percentage of reports related to flaky tests (Column $P$) computed as $F / T \times 100\%$.}
    \label{tab:repo}
    \begin{tabular}{@{}lllrrr@{}}    
    \toprule
    \emph{Platform} & \emph{Repository} & \emph{Language} & $T$ & $F$ & $P$ \\
    \midrule
    Qiskit & qiskit    & Python & 4,533  & 29 & 0.55\% \\ 
    Qiskit & qiskit-aer    & Python & 766  & 3 & 0.39\% \\ 
     Qiskit & qiskit-ibm-runtime & Python & 849 & 3 & 0.35\% \\
    Qiskit & qiskit-ibm-provider & Python & 341  & 8 & 0.23\% \\
    Qiskit Community & qiskit-nature & Python & 385 & 1 & 0.26\% \\ 
    Qiskit Community & qiskit-experiments & Python & 389 & 3 & 0.77\% \\
    Qiskit Community & qiskit-machine-learning & Python & 306 & 4 & 1.30\% \\
    Microsoft & azure-quantum-python & Python & 89 & 3 & 3.37\% \\
    Microsoft & QuantumLibraries & Q\# & 137 & 4 & 2.91\% \\
    Microsoft & Quantum & Q\# & 111 & 4 & 3.60\% \\
    TensorFlow & quantum & Python & 306 & 1 & 0.32\% \\
    NetKet & netket & Python & 416 & 7 & 1.68\% \\
    \midrule
    Total & & & 8,628 & 71 &   \\
    \bottomrule
    \end{tabular}
\end{table*}

To answer \textbf{RQ1} and detect more flaky reports systematically and automatically, we employed embedding transformers to represent GitHub IRs and PRs and measured the cosine similarity with previously identified flaky tests.

Based on the Hugging Face leaderboard~\cite{muennighoff2022mteb}, we selected three top-performing (at the time of experiment design) embeddings on generic tasks. We utilized the pre-trained \texttt{mixedbread-ai/mxbai-embed-large-v1} transformer~\cite{emb2024mxbai,li2023angle} from the Hugging Face library~\cite{wolf2019huggingface} to extract contextual embeddings\footnote{The embeddings were derived from the penultimate layer of the model, ensuring that they captured the nuanced features of the text.} of the GitHub IRs and PRs. We also evaluated other transformers, such as \texttt{SFR-Mistral} ~\cite{SFRAIResearch2024} and \texttt{e5-mistral-7b-instruct} transformers ~\cite{wang2024improving,wang2022text}, using $k$-means clustering~\cite{macqueen1967some} to assess their effectiveness in distinguishing flaky and non-flaky test cases. Our analysis showed that the model of \texttt{mixedbread-ai/mxbai-embed-large-v1} provided the most distinct and separable representation for this task.

Using this model, we generated embeddings from the tokenized text of all scraped GitHub IRs and PRs from the 12 repositories, as well as from the known flaky cases in our source dataset. We calculated cosine similarity scores between the embeddings and ranked the issues based on their similarity to known flaky cases. Excluding identical matches from our initial flaky set, at least two authors cross-examined the top-ranked issues, associated PRs, and code commits to determine if they were related to flaky tests and to establish their cause categories.

In the first iteration, we identified 15 new flaky tests from the top-ranked cosine similarity scores. We then augmented the original dataset with these new cases, performed manual cross-examination, and repeated the process, resulting in the detection of an additional 10 flaky tests in the second iteration. In total,\textit{ we identified 25 new flaky tests across the 12 repositories, increasing the original dataset size by 54\%.}

In Table~\ref{tab:repo}, we summarize the dataset, which contains 71 flaky tests collected from 12 open-source quantum software repositories.\footnote{We retain the original platform naming from our initial study for consistency; however, some Qiskit repositories have since been restructured (e.g., \texttt{qiskit-ibm-provider} has been merged into \texttt{qiskit-ibm-runtime}).} 
Overall, we identify $71$ flaky tests from $8{,}628$ closed IRs and their associated PRs, corresponding to an observed flakiness rate of approximately $0.82\%$ ($71/8{,}628$). This rate may appear lower than what has been reported in large-scale industrial studies of classical software; a plausible explanation is that many quantum software projects are comparatively younger, smaller in codebase size, and less test-intensive, and that flakiness may also be underreported or described using inconsistent terminology. Nevertheless, the practical impact of quantum flakiness can be substantial: reproducing intermittent failures may require repeated executions, and diagnosis often demands interdisciplinary expertise spanning software engineering and quantum algorithms.

As can be seen from Table~\ref{tab:repo}, \texttt{qiskit} has the largest issue volume ($T{=}4{,}533$) yet a modest flaky prevalence ($P{=}0.55\%$). Several other Qiskit-related repositories (in Python) fall below $1\%$ (e.g., \texttt{qiskit-aer} is $0.39\%$), which may reflect stronger testing infrastructures. By contrast, smaller repositories can show higher $P$ values. For instance, Microsoft \texttt{Quantum} ($T{=}111$) shows $P{=}3.60\%$, which is based on Q\#, suggesting that language, toolchain, hardware interface factors may influence the flakiness.

\textbf{Note: }Because Table~\ref{tab:repo} is derived from developer-reported IRs/PRs, it likely underestimates the true prevalence of quantum flakiness, i.e., some flaky behaviors may never be reported or reported using inconsistent terminology. This observation motivates more efficient and effective automated detection methods that can proactively identify flaky-test-related IRs/PRs.

\begin{table*}[t]
\centering
\large
\caption{Count of cause categories and fix patterns based on flaky test reports.}
\label{tab:cause}
\resizebox{\textwidth}{!}{\begin{tabular}{@{}l|rrrrrrrr|rr@{}}
\toprule
                     & \multicolumn{8}{c|}{Fix pattern}                                                                                                          &              &  \\ \cmidrule(lr){2-9}
Cause category       & Fix Seed & Alter Software Env. & Make Single Thread & Adjust Tolerance & Add Exception Handler & Synchronize & Use Keys for Order & Others & Grand Total & Percentage    \\ \midrule
Randomness           & 12        &                     &                    &                  &                       &             &                    & 2      & 14          & 19.2\%  \\
Software Env.        &          & 5                   &                    &                  &                       &             &                    & 3      & 8           & 11.0\%  \\
Multi-Threading      &          &                     & 4                  &                  &                       &             &                    & 6      & 10           & 13.7\%  \\
Floating Point Ops.  &          &                     &                    & 5                &                       &             &                    & 2      & 7           & 9.6\%  \\
Visualization        &          &                     &                    &                  &                       &             &                    & 4      & 4           & 5.5\%   \\
Unhandled Exception  &          &                     &                    &                  & 4                     &             &                    &        & 4           & 5.5\%   \\
Network              &          &                     &                    &                  &                       & 1           &                    &  5      & 6           & 8.2\%   \\
Unordered Collection &          &                     &                    &                  &                       &             & 1                  &        & 1           & 1.4\%   \\
Others               &          &                     &                    &                  &                       &             &                    & 12      & 12           & 16.4\%  \\
Unknown              &          &                     &                    &                  &                       &             &                    & 7      & 7           & 9.6\%   \\ \midrule
Grand Total          & 12        & 5                  & 4                  & 5               & 4                     & 1           & 1                  & 41     & 73          & 100\% \\
Percentage           & 16.4\%     & 6.8\%                & 5.5\%                & 6.8\%              & 5.5\%                   & 1.4\%         & 1.4\%                & 56.2\%   & 100\%       &      \\ \bottomrule
                     
\end{tabular}}
\end{table*}

\subsection{Root Cause Analysis and Common Fixes}

To answer \textbf{RQ2}, Table~\ref{tab:cause} provides the root cause analysis and common fix patterns for the 71 quantum flaky tests. Note that Table~\ref{tab:cause} contains 73 observations because 2 flaky tests have multiple labels (e.g., a flaky test can be caused by both randomness and floating point operations). As can be seen in the table, the most common root cause is ``Randomness'' (i.e., 19.2\% of 73 cases), and its most common fix is to ``Fix Seed'' (i.e., 16.4\% in all fixes). Compared to classical flakiness in~\cite{luo2014empirical}, the leading causes are async waits and concurrency. 

Our findings are closer to those focusing on probabilistic programming and machine learning software, where 60\% of flaky tests may be caused by randomness~\cite{dutta2020detecting}. As PRNGs are heavily used in quantum programming, randomness contributes significantly to flaky test reports in quantum programs.
\subsubsection{Randomness/Fix seed}
As mentioned earlier,  the use of PRNGs is the most common (19.2\%) cause of reports of flakiness in quantum programs. The PRNGs result in flaky test since they produce different output from run to run. 
To illustrate the randomness, we show an example in Listing~\ref{lst:seed} (based on issue report \#5217 in \texttt{qiskit}). The logic of the test \texttt{test\_append\_circuit}, which checks if appending quantum circuits work properly, is correct. However, Lines 3 and 6 cause flakiness in the test because the function \texttt{random\_circuit} generates a random quantum circuit using a randomly selected seed (by default). Thus, the test fails occasionally due to randomness.

\begin{lstlisting}[language=Python, label={lst:seed}, caption=An example of a PRNG-related flaky test in the code of a test case.]
def test_append_circuit(self, num_qubits):
    ...
    first_circuit = random_circuit(num_qubits[0], depth)
    ...
    for num in num_qubits[1:]:
        circuit = random_circuit(num, depth)
\end{lstlisting}

12 out of 14 randomness-related flaky test reports (i.e., 16.4\% of all flaky test reports) are fixed by setting a random seed value to a constant (or in combination with a reduced convergence threshold as shown in pull request \#8820 of \texttt{qiskit}). For example, to avoid flakiness in Listing~\ref{lst:seed}, Listing~\ref{lst:fix-seed} (pull request \#5599 in \texttt{qiskit}) shows a solution that replaces the default non-constant random seed value with a fixed seed value. More examples are provided in our prior work~\cite{zhang2023identifying}.
\begin{lstlisting}[language=Python, label={lst:fix-seed}, caption=An example of fixed seed for Listing~\ref{lst:seed}.]
- first_circuit = random_circuit(num_qubits[0], depth)
+ first_circuit = random_circuit(num_qubits[0], depth, seed=4200)
...
- circuit = random_circuit(num, depth)
+ circuit = random_circuit(num, depth, seed=4200)
\end{lstlisting}

\subsubsection{Software Environment/Alter Software Environment}
This category includes flaky tests caused by specific software or library dependency issues. For example, pull request \#1369 in \texttt{netket} discusses a flaky test observed only in the GitHub Actions Python 3.10 environment. The pull request starts with the following comment. 

\begin{displayquote}
``No idea why, but on the GitHub runner python 3.10 this test keeps failing. I never reproduced locally so I think it's not a real failure. Simplyfying [\textit{sic}] the test to avoid.''
\end{displayquote}

A common way to fix this issue is to alter the software environment as it is shown in Listing~\ref{lst:fix-env}. 

\begin{lstlisting}[language=Python, label={lst:fix-env}, caption=An example of fixing library dependency.]
- numpy!=1.19
\end{lstlisting}

\subsubsection{Multi-Threading/Make Single Thread}
This category of flaky tests is caused by multi-threading issues, e.g., concurrency and overload. As an example, issue report \#5904 in \texttt{qiskit} describes a flaky test caused by address collisions due to parallel builds. The code in this example can be seen in Listing~\ref{lst:multi-thread}, showing that \texttt{sphinx-build} generates documentation in parallel over $N$ processes, where $N$ is the number of CPUs (i.e., by the argument \texttt{-j auto}). While the parallelization improves the throughput, it causes an error when multiple jobs of \texttt{sphinx-build} compete against each other. A fix pattern for this cause is given in Listing~\ref{lst:fix-thread}.
\begin{lstlisting}[language=Ini, label={lst:multi-thread},
caption=An example of a flaky test related to multi-threading.]
commands = sphinx-build -W -b html -j auto docs/ docs/_build/html {posargs}
\end{lstlisting}

\begin{lstlisting}[language=Python, label={lst:fix-thread},
caption=The fix for the flaky test in Listing~\ref{lst:multi-thread}.]
- sphinx-build -W -b html -j auto docs/ docs/_build/html {posargs}
+ sphinx-build -W -b html docs/ docs/_build/html {posargs}
\end{lstlisting}
\subsubsection{Floating Point Operations/Adjust Tolerance}
We define floating-point-related flaky tests as those whose nondeterministic behavior is caused by numerical precision problems, such as round-off errors. For example, in \texttt{netket} pull request  \#1147 (Listing~\ref{lst:netkit_assert}), the hard-coded tolerance value of $10^{-5}$ (shown in Line 5) causes a test case to fail intermittently.

\begin{lstlisting}[language=Python, label={lst:netkit_assert},
  caption=An example of a flaky test related to floating point operations.]
def test_vmc_functions():
    ha, sx, ma, sampler, driver = _setup_vmc()
    driver.advance(500)
    assert driver.energy.mean == \ 
        approx(ma.expect(ha).mean, abs=1e-5)
\end{lstlisting}

For flaky tests caused by floating-point operations, adjusting numerical tolerances can mitigate the issue. Five of the seven reports (9.6\% of all reports) were resolved using this approach. For example, one can increase the tolerance manually, e.g., by doubling the tolerance value as shown in Listing~\ref{lst:netkit_assert}. However, hard-coded loose tolerances may result in increased false negative test results, affecting the correctness of the program.

\begin{lstlisting}[language=Python, label={lst:fix_netkit_assert}, caption=An example of tolerance increase for Listing~\ref{lst:netkit_assert}.]
- assert driver.energy.mean == approx(ma.expect(ha).mean, abs=1e-5)
+ tol = driver.energy.error_of_mean * 5
+ assert driver.energy.mean == approx(ma.expect(ha).mean, abs=tol)
\end{lstlisting}

\subsubsection{Visualization}

This group of flaky tests is related to image generation. For example, \texttt{qiskit} has a test manager that schedules visual tests sequentially and allows these tests to communicate with one another. Issue \#3283 in \texttt{qiskit} reports a flaky test in the visualizer of the test manager due to out-of-date reference indexes.

As noted by Zhang et al.~\cite{zhang2023identifying}, no recurring fix pattern can be identified. Each of the four reports (all associated with \texttt{qiskit}) require a distinct solution.

\subsubsection{Unhandled Exception/Add Exception Handler}

This group of flaky tests is caused by the code that does not appropriately handle exceptions. For example, Listing~\ref{lst:unhandled} shows the stack trace in the issue report of \#398 in \texttt{Microsoft/QuantumLibraries}. In this case, the testing can occasionally fail whenever the ``number'' does not fall into the expected range.

\begin{lstlisting}[language=bash, label={lst:unhandled}, caption=An example of an unhandled exception.]
Unhandled exception.
Microsoft.Quantum.Simulation.Core.ExecutionFailException: "number" must be between 0 and 2^3 - 1, but was -1.
\end{lstlisting}

Flaky tests caused by unhandled exceptions can be mitigated by adding an exception handler or removing the exception (all four reports, or 5.5\% of all the reports, are fixed this way). For example, developers add an \texttt{if} condition to remove any unexpected negative coefficients in pull request \#399 of \texttt{Microsoft/QuantumLibraries} (see Listing~\ref{lst:fix-exception}). 

\begin{lstlisting}[language=Python, label={lst:fix-exception},
caption=The fix for the flaky test in Listing~\ref{lst:unhandled}.]
- ApplyXorInPlace(keepCoeff[idx], keepCoeffRegister);
+ if (keepCoeff[idx] >= 0) {
+   ApplyXorInPlace(keepCoeff[idx], keepCoeffRegister);
+ }
\end{lstlisting}

\subsubsection{Network/Synchronize}
Flakiness in this category is due to network-related issues, such as unstable networks or servers. For example, issue \#584 of \texttt{qiskit-ibm-runtime} reports a flaky test due to timeouts and socket connection problems. As shown in Listing~\ref{lst:network}, \texttt{test\_websocket\_proxy} fails because jobs have been completed before \texttt{websocket} connection can be established. A method to fix this cause is given in Listing~\ref{lst:fix-network}.

\begin{lstlisting}[language=bash, label={lst:network}, caption=An example of network-related flaky test.]
FAIL: test_websocket_proxy (test.integration.test_results.TestIntegrationResults) 
    (service=<QiskitRuntimeService>)
\end{lstlisting}

We observe five (i.e., 8.2\% of all the reports) network-related flaky test report, i.e., issue \#584 and pull request \#588 in \texttt{qiskit-ibm-runtime}. Here, tests for callback functions are dependent on socket tests. However, sometimes those callback tests finish before the socket tests, which causes flakiness. Developers increase callback test iterations so that socket tests have sufficient time to finish first (see Listing~\ref{lst:fix-network}). This may be a suboptimal solution as the number of iterations is hard-coded. The ideal solution would be to synchronize callback tests with the completion of socket tests.

\begin{lstlisting}[language=Python, label={lst:fix-network},
caption=The fix for the flaky test in Listing~\ref{lst:network}.]
- job = self._run_program(service, iterations=1, callback=result_callback)
+ job = self._run_program(service, iterations=10, callback=result_callback)
\end{lstlisting}

\subsubsection{Unordered Collection/Use Keys for Order}

This category of flaky tests is Python-specific (although, hypothetically, it may appear in other languages). Dictionaries (hash maps or hash tables) are implemented as an unordered collection from Python 3.3 to 3.7~\cite{gruber2021empirical}. When tests have order dependencies, test outcomes can become non-deterministic, which leads to flakiness. For example, pull request \#8627 in \texttt{qiskit} reveals a flaky test that uses insertion order to compare two dictionaries, hence non-deterministic results.

For flaky tests caused by unordered Python dictionaries, developers can use key values for ordering instead of using the insertion order. 
In Python 3.8 and later, dictionaries are order-preserved, so upgrading the Python environment can solve the problem. However, Python upgrades can cause dependency issues. One flaky test report (i.e., 1.4\% of all reports) had this cause and fix.

\section{Automated Quantum Flakiness Detection}\label{sec:automation}

This section presents our automated pipeline for identifying quantum flaky-test-related IRs/PRs and diagnosing their likely root causes. We first describe how we construct model inputs by combining textual evidence from IR/PR content with optional code context extracted from the affected files. We then detail the prompting and inference configurations used to evaluate LLMs under different context settings, enabling systematic comparison of detection performance and root-cause identification accuracy.

\subsection{Input preparation}\label{subsec:dataset_prep}

To create a balanced dataset, we also collected non-flaky tests from GitHub reports during the cosine similarity analysis. Test cases with a cosine similarity score of less than 0.5 were labeled as non-flaky. Additionally, we included instances that our method incorrectly labeled as flaky to further challenge the classification task. In total, we compiled 71 non-flaky cases to match the size of our expanded flaky dataset.

We extracted descriptions and comments from IRs and PRs for each issue observation in the dataset, recorded code differences, and noted affected files before and after fixes. Method-level code changes were also extracted for a more concentrated analysis.

Manual verification ensured that the extracted artifacts matched the identified IRs and PRs. The dataset was organized into ``full'' and ``method'' directories, each further divided into ``flaky'' and ``non-flaky'' sections.

\subsection{Detecting Flaky Tests with LLMs}

\begin{figure*}
    \centering
    \includegraphics[width=1\linewidth]{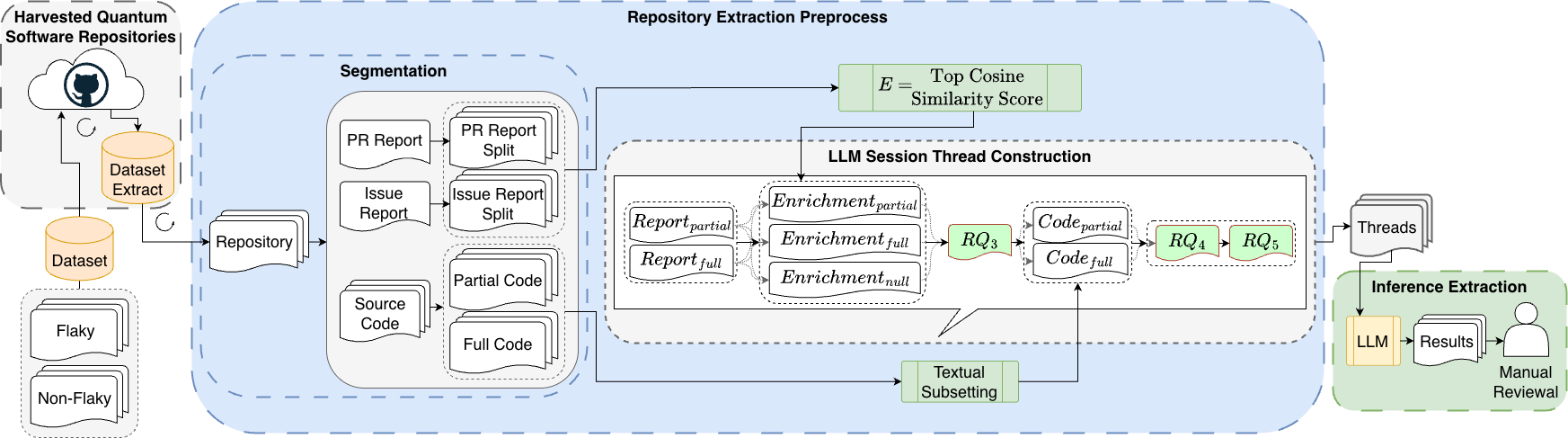}
    \caption{Automated quantum flakiness detection and root cause analysis pipeline architecture.}
    \label{fig:diagram}
\end{figure*}

LLMs are large-scale pretrained models developed through substantial upfront investments in data, compute, and expertise have become transformative tools across diverse scientific domains, owing to their ability to be efficiently adapted to a wide range of downstream tasks. By leveraging this upfront training effort, LLMs demonstrate remarkable generalization and adaptability with minimal task-specific tuning, enabling rapid progress even in data-scarce scientific settings~\cite{trivsovic2025rapid, sivaloganathan2025mapping}.

In the context of quantum computing, where datasets are often limited, expensive to generate, and highly specialized, LLMs offer a particularly compelling paradigm: pretrained vision and language models can be repurposed to extract structure, patterns, and abstractions from quantum data without the prohibitive cost of training bespoke neural networks from scratch. The rapid adoption of LLMs in science—particularly open-weight vision and language models in computer science and engineering—underscores their broad utility and cost-effectiveness for research applications where training artificial neural networks is impractical~\cite{trivsovic2025rapid}.

This study focuses on a selected set of LLMs, chosen for their suitability to our objectives: 1) source code interpretability, augmentation, and analysis as a linguistic modality; 2) the limited dataset size under study; 3) their relatively low cost; and 4) their effectiveness compared to other existing solutions~\cite{akli2023flakycat, fatima2024flakyfix}. Compared to other solutions using few-shot learning approaches on Java~\cite{akli2023flakycat, fatima2024flakyfix}, we focus on Qiskit and NetKet (at method level), which are based on assembly-like language and wrapped in Python, and our experiments demonstrate an automated LLM-based pipeline (see Fig.~\ref{fig:diagram}) that efficiently gathers resources, configures inputs, and classifies bugs by streamlining the standard software development workflow with GitHub as the version control system. This pipeline simulates a typical software engineering process for bug resolution.

\subsection{LLM Inference Configuration}\label{subsec:configuration}

Leveraging the extracted codebase data (discussed in section \ref{subsec:dataset_prep}), we dynamically construct prompts to address our research questions. The final prompts are provided in the supplementary material~\cite{sivaloganathan_2026_18642687}. We welcome community input for further refinement.

To explore our research questions and assess how context arguments correlates with the answers, we designed the following experiments.

For \textbf{RQ3}, which aims to classify whether a particular IR (or PR, if no issue is associated with it) is flaky or non-flaky, we tested two levels of context: $R_p$ (partial), which includes only the initial IR (or PR) description, and $R_f$ (full), which includes the description along with all associated comments.

For \textbf{RQ4}, we expanded the context for the language model by adding the code involved in the PR before the fix was applied, also at two levels: $C_p$ (partial), which includes the method-level code, and $C_f$ (full), which provides the complete code listing.

By combining the context levels from RQ3 and RQ4, we generated four experimental conditions: $\{R_p, R_f\} \times \{C_p, C_f\}$. These conditions range from $(R_p, C_p)$, which uses only the description and method-level code, to $(R_f, C_f)$, which includes the description with comments and the full code listing.

For \textbf{RQ5}, the amount of information provided did not change; we simply followed up by querying which specific root cause a particular flaky test relates to, using the nine classes defined by~\cite{zhang2023identifying}: ``Randomness (PRNG)'', ``Floating Point Operations'', ``Software Environment'', ``Multi-threading'', ``Visualization'', ``Unhandled Exceptions'', ``Network'', ``Unordered Collection'', or ``Others''. We find that we do not need to alter these classes for our extended dataset.

We utilized LangChain~\cite{LangChai97:online}, an integration framework that abstracts prompt templating, retrieval strategies, and chaining, allowing us to manage conversational memory effectively. This setup enabled us to simulate a Developer-to-AI interaction, appending additional context to study its impact on the LLM's reasoning.

\subsection{Context Enrichment Strategies}
To further assess pipeline performance tuning options, we executed three context enrichment strategies. Cosine similarity rankings on the embeddings of our data derived from \texttt{mixedbread-ai/mxbai-embed-large-v1} model were used against the same dataset to identify the highest scored issue reports for $R_p$ ($E_p$) and $R_f$ ($E_f$) respectively. The reports are then used as RQ3 imputations for a guiding example. The additional enrichments then updates the research design conditions to: $\{R_p, R_f\} \times \{C_p, C_f\} \times \{E_\emptyset
, E_p, E_f\}$ where $E_{\emptyset}$ represents null context. The associated $C_p$ and $C_f$ for each $E_p$ and $E_f$ were also added to RQ3 as an additional example for few-shot prompting.

\subsection{LLMs under study} 
To assess the performance of a diverse set of LLMs, we study two large open-source models, namely Meta \texttt{llama-3.1-70b} and \texttt{llama-3.1-405b}~\cite{dubey2024LLaMA}, as well as five closed-source models, namely OpenAI \texttt{gpt-4o}, \texttt{gpt-4o-mini}, and \texttt{gpt-4.1-mini}~\cite{openai2024gpt4omini,openai2024gpt4o,openai2025gpt41} and Google \texttt{gemini-2.5-flash}~\cite{google2025gemini25flash}, and Anthropic \texttt{claude-4-haiku}~\cite{anthropic2025claudehaiku}. All instruct models used were accessed remotely via serverless APIs of OpenAI~\cite{Overview51:online}, Anthropic~\cite{AnthropicAPIOverview:online}, and Google VertexAI~\cite{VertexAI96:online}.\footnote{We also fine-tuned CodeBERT~\cite{feng2020codebert} for flaky test detection using a few-shot learning approach. While training was successful, the model failed to generalize effectively for the test set when applied to GitHub IRs and PRs.}

\section{Results and Analysis}\label{sec:results}

\begin{table*}[htbp]
\centering
\caption{Model Performances}
\label{tab:results}
\resizebox{0.65\textwidth}{0.48\textheight}{% % This scales the table to fit the page width
\begin{tabular}{@{}ll ccccccccc@{}} % Corrected to 11 columns
\toprule
\multirow{2}{*}{\textbf{Model}} &
\multirow{2}{*}{\textbf{Context}} &
\multicolumn{2}{c}{\textbf{F1}} &
\multicolumn{2}{c}{\textbf{MCC}} &
\multicolumn{2}{c}{\textbf{Recall}} &
\multicolumn{3}{c}{\textbf{Total Observations}} \\
\cmidrule(lr){3-4} \cmidrule(lr){5-6} \cmidrule(lr){7-8} \cmidrule(lr){9-11}
& & RQ3 & RQ5 & RQ3 & RQ5 & RQ3 & RQ4 & RQ3 & RQ4 & RQ5 \\
\midrule

% GPT-4o
\multirow{12}{*}{\textbf{\texttt{gpt-4o-2024-11-20}}}
  & $\{R_p, C_p\}$      & \multirow{2}{*}{0.8592} & 0.5069 & \multirow{2}{*}{0.7183} & 0.5187 & \multirow{2}{*}{0.8592} & 0.8182          & \multirow{12}{*}{142} & 44 & 44 \\
  & $\{R_p, C_f\}$      &                         & 0.4586          &                         & 0.4386          &                         & 0.6406         &  & 64 & 64 \\
  & $\{R_f, C_p\}$      & \multirow{2}{*}{0.8649} & 0.4968          & \multirow{2}{*}{0.7209} & 0.5734          & \multirow{2}{*}{0.9014} & 0.9211 &  & 38 & 38 \\
  & $\{R_f, C_f\}$      &                         & 0.5481          &                         & 0.5802          &                         & 0.8393          &  & 56 & 56 \\
\cmidrule(l){2-8} \cmidrule(l){10-10}
  & $E_p\{R_p, C_p\}$   & \multirow{2}{*}{0.8182} & 0.4137 & \multirow{2}{*}{0.6145} & 0.4682 & \multirow{2}{*}{0.8732} & 0.8182 &  & 44 & 44 \\
  & $E_p\{R_p, C_f\}$   &                         & 0.4098 &                         & 0.4360 &                         & 0.7969 &  & 64 & 64 \\
  & $E_p\{R_f, C_p\}$   & \multirow{2}{*}{0.8456} & 0.5262 & \multirow{2}{*}{0.6794} & 0.6006 & \multirow{2}{*}{0.8873} & {0.8684} &  & 38 & 38 \\
  & $E_p\{R_f, C_f\}$   &                         & 0.5371 &                         & 0.5863 &                         & 0.8571  &  & 56 & 56 \\
\cmidrule(l){2-8} \cmidrule(l){10-10}
  & $E_f\{R_p, C_p\}$   & \multirow{2}{*}{0.7898} & 0.4398 & \multirow{2}{*}{0.5476} & 0.5043 & \multirow{2}{*}{0.8732} & 0.8636 &  & 44 & 44 \\
  & $E_f\{R_p, C_f\}$   &                         & 0.4712 &                         & 0.5030 &                         & 0.8065 &  & 62$^*$ & 62$^*$ \\
  & $E_f\{R_f, C_p\}$   & \multirow{2}{*}{0.8421} & 0.5149 & \multirow{2}{*}{0.6686} & 0.5763 & \multirow{2}{*}{0.9014} & 0.8947 &  & 38 & 38 \\
  & $E_f\{R_f, C_f\}$   &                         & 0.5494 &                         & 0.6003 &                         & 0.8545 &  & 55$^*$ & 55$^*$ \\

\midrule

% GPT-4o-mini - updated
\multirow{12}{*}{\textbf{\texttt{gpt-4o-mini-2024-07-18}}}
  & $\{R_p, C_p\}$      & \multirow{2}{*}{0.8421} & 0.4845 & \multirow{2}{*}{0.7100} & 0.5127 & \multirow{2}{*}{0.7887} & 0.5682 & \multirow{12}{*}{142} & 44 & 44 \\
  & $\{R_p, C_f\}$      &                         & 0.3870 &                         & 0.4022 &                         & 0.4531 &  & 64 & 64 \\
  & $\{R_f, C_p\}$      & \multirow{2}{*}{0.8794} & 0.5033 & \multirow{2}{*}{0.7606} & 0.5703 & \multirow{2}{*}{0.8732} & 0.6579 &  & 38 & 38 \\
  & $\{R_f, C_f\}$      &                         & 0.4858 &                         & 0.5020 &                         & 0.6429 &  & 56 & 56 \\
\cmidrule(l){2-8} \cmidrule(l){10-10}
  & $E_p\{R_p, C_p\}$   & \multirow{2}{*}{0.8552} & 0.4532 & \multirow{2}{*}{0.7049} & 0.5537 & \multirow{2}{*}{0.8732} & 0.5455 &  & 44 & 44 \\
  & $E_p\{R_p, C_f\}$   &                         & 0.3845 &                         & 0.4178 &                         & 0.5469 &  & 64 & 64 \\
  & $E_p\{R_f, C_p\}$   & \multirow{2}{*}{0.9041} & 0.4351 & \multirow{2}{*}{0.8041} & 0.5069 & \multirow{2}{*}{0.9296} & 0.6316 &  & 38 & 38 \\
  & $E_p\{R_f, C_f\}$   &                         & 0.4529 &                         & 0.4749 &                         & 0.5714 &  & 56 & 56 \\
\cmidrule(l){2-8} \cmidrule(l){10-10}
  & $E_f\{R_p, C_p\}$   & \multirow{2}{*}{0.8889} & 0.4199 & \multirow{2}{*}{0.7750} & 0.4964 & \multirow{2}{*}{0.9014} & 0.5909 &  & 44 & 44 \\
  & $E_f\{R_p, C_f\}$   &                         & 0.3920 &                         & 0.4183 &                         & 0.4688 &  & 64 & 64 \\
  & $E_f\{R_f, C_p\}$   & \multirow{2}{*}{0.9103} & 0.4551 & \multirow{2}{*}{0.8176} & 0.5391 & \multirow{2}{*}{0.9296} & 0.5789 &  & 38 & 38 \\
  & $E_f\{R_f, C_f\}$   &                         & 0.4995 &                         & 0.5227 &                         & 0.5714 &  & 56 & 56 \\

\midrule

% GPT-4.1-mini
\multirow{12}{*}{\textbf{\texttt{gpt-4.1-mini-2025-04-14}}}
  & $\{R_p, C_p\}$      & \multirow{2}{*}{0.8462} & 0.4363 & \multirow{2}{*}{0.7288} & 0.4314 & \multirow{2}{*}{0.7746} & 0.7500 & \multirow{12}{*}{142} & 44 & 44 \\
  & $\{R_p, C_f\}$      &                         & 0.4912 &                         & 0.4667 &                         & 0.7500 &  & 64 & 64 \\
  & $\{R_f, C_p\}$      & \multirow{2}{*}{0.8857} & 0.4001 & \multirow{2}{*}{0.7750} & 0.4401 & \multirow{2}{*}{0.7746} & 0.8947 &  & 38 & 38 \\
  & $\{R_f, C_f\}$      &                         & {\textbf{0.6457}} &                         & {\textbf{0.6375}} &                         & 0.8571 &  & 56 & 56 \\
\cmidrule(l){2-8} \cmidrule(l){10-10}
  & $E_p\{R_p, C_p\}$   & \multirow{2}{*}{0.8397} & 0.4313 & \multirow{2}{*}{0.7128} & 0.4633 & \multirow{2}{*}{0.7746} & 0.7727 &  & 44 & 44 \\
  & $E_p\{R_p, C_f\}$   &                         & 0.5400 &                         & 0.5416 &                         & 0.7188 &  & 64 & 64 \\
  & $E_p\{R_f, C_p\}$   & \multirow{2}{*}{0.8593} & 0.4978 & \multirow{2}{*}{0.7360} & 0.5395 & \multirow{2}{*}{0.8169} & 0.7895 &  & 38 & 38 \\
  & $E_p\{R_f, C_f\}$   &                         & 0.6072 &                         & 0.5755 &                         & 0.8036 &  & 56 & 56 \\
\cmidrule(l){2-8} \cmidrule(l){10-10}
  & $E_f\{R_p, C_p\}$   & \multirow{2}{*}{0.8244} & 0.4477 & \multirow{2}{*}{0.6843} & 0.4278 & \multirow{2}{*}{0.7606} & 0.7500 &  & 44 & 44 \\
  & $E_f\{R_p, C_f\}$   &                         & 0.4685 &                         & 0.4574 &                         & 0.7656 &  & 64 & 63$^*$ \\
  & $E_f\{R_f, C_p\}$   & \multirow{2}{*}{0.8467} & 0.5275 & \multirow{2}{*}{0.7060} & 0.5325 & \multirow{2}{*}{0.8169} & 0.8158 &  & 38 & 38 \\
  & $E_f\{R_f, C_f\}$   &                         & 0.5933 &                         & 0.5767 &                         & 0.7679 &  & 56 & 56 \\

\midrule

% LLaMA-70B-Instruct
\multirow{12}{*}{\textbf{\texttt{llama3.1:70b-instruct}}}
  & $\{R_p, C_p\}$      & \multirow{2}{*}{0.8264} & 0.5024 & \multirow{2}{*}{0.7372} & 0.5038 & \multirow{2}{*}{0.7042} & 0.5455 & \multirow{12}{*}{142} & 44 & 44 \\
  & $\{R_p, C_f\}$      &                         & 0.4642 &                         & 0.4590 &                         & 0.5156 &  & 64 & 64 \\
  & $\{R_f, C_p\}$      & \multirow{2}{*}{0.8067} & 0.4415 & \multirow{2}{*}{0.7146} & 0.4677 & \multirow{2}{*}{0.6761} & 0.5789 &  & 38 & 38 \\
  & $\{R_f, C_f\}$      &                         & 0.5429 &                         & 0.5729 &                         & 0.5179 &  & 56 & 56 \\
\cmidrule(l){2-8} \cmidrule(l){10-10}
  & $E_p\{R_p, C_p\}$   & \multirow{2}{*}{0.8000} & 0.4829 & \multirow{2}{*}{0.6962} & 0.5106 & \multirow{2}{*}{0.6761} & 0.4091 &  & 44 & 44 \\
  & $E_p\{R_p, C_f\}$   &                         & 0.4276 &                         & 0.4315 &                         & 0.4219 &  & 64 & 64 \\
  & $E_p\{R_f, C_p\}$   & \multirow{2}{*}{0.7899} & 0.4668 & \multirow{2}{*}{0.6848} & 0.4991 & \multirow{2}{*}{0.6620} & 0.6316 &  & 38 & 38 \\
  & $E_p\{R_f, C_f\}$   &                         & 0.5178 &                         & 0.5532 &                         & 0.4821 &  & 56 & 56 \\
\cmidrule(l){2-8} \cmidrule(l){10-10}
  & $E_f\{R_p, C_p\}$   & \multirow{2}{*}{0.8067} & 0.4559 & \multirow{2}{*}{0.7146} & 0.4487 & \multirow{2}{*}{0.6761} & 0.4091 &  & 44 & 44 \\
  & $E_f\{R_p, C_f\}$   &                         & 0.4223 &                         & 0.4106 &                         & 0.4127 &  & 63$^*$ & 64 \\
  & $E_f\{R_f, C_p\}$   & \multirow{2}{*}{0.8000} & 0.4210 & \multirow{2}{*}{0.6962} & 0.4370 & \multirow{2}{*}{0.6761} & 0.6053 &  & 38 & 38 \\
  & $E_f\{R_f, C_f\}$   &                         & 0.5265 &                         & 0.5572 &                         & 0.4107 &  & 56 & 56 \\

\midrule

  % LLaMA-405B-Instruct - updated
\multirow{12}{*}{\textbf{\texttt{llama3.1:405b-instruct}}}
  & $\{R_p, C_p\}$      & \multirow{2}{*}{0.8333} & 0.4660 & \multirow{2}{*}{0.6971} & 0.4818 & \multirow{2}{*}{0.7746} & 0.7727 & \multirow{12}{*}{142} & 44 & 44 \\
  & $\{R_p, C_f\}$      &                         & 0.3984 &                         & 0.3766 &                          & 0.7031 &  & 64 & 64 \\
  & $\{R_f, C_p\}$      & \multirow{2}{*}{0.8759} & 0.4514 & \multirow{2}{*}{0.7625} & 0.5077 & \multirow{2}{*}{0.8451} & 0.7895 &  & 38 & 38 \\
  & $\{R_f, C_f\}$      &                         & 0.5077 &                         & 0.5356 &                          & 0.7321 &  & 56 & 56 \\
\cmidrule(l){2-8} \cmidrule(l){10-10}
  & $E_p\{R_p, C_p\}$   & \multirow{2}{*}{0.8182} & 0.4958 & \multirow{2}{*}{0.6686} & 0.5437 & \multirow{2}{*}{0.7606} & 0.6591 &  & 44 & 44 \\
  & $E_p\{R_p, C_f\}$   &                         & 0.4279 &                         & 0.4191 &                          & 0.6406 &  & 64 & 64 \\
  & $E_p\{R_f, C_p\}$   & \multirow{2}{*}{0.8397} & 0.4743 & \multirow{2}{*}{0.7128} & 0.5380 & \multirow{2}{*}{0.7746} & 0.7632 &  & 38 & 38 \\
  & $E_p\{R_f, C_f\}$   &                         & 0.4983 &                         & 0.5152 &                          & 0.6607 &  & 56 & 56 \\
\cmidrule(l){2-8} \cmidrule(l){10-10}
  & $E_f\{R_p, C_p\}$   & \multirow{2}{*}{0.8125} & 0.5001 & \multirow{2}{*}{0.6752} & 0.5456 & \multirow{2}{*}{0.7324} & 0.6818 &  & 44 & 44 \\
  & $E_f\{R_p, C_f\}$   &                         & 0.4193 &                         & 0.4014 &                         & 0.5938 &  & 64 & 63 \\
  & $E_f\{R_f, C_p\}$   & \multirow{2}{*}{0.8571} & 0.4743 & \multirow{2}{*}{0.7662} & 0.5380 & \multirow{2}{*}{0.7606} & 0.7368 &  & 38 & 38 \\
  & $E_f\{R_f, C_f\}$   &                         & 0.5158 &                         & 0.5379 &                         & 0.6250 &  & 56 & 56 \\

\midrule

% Gemini 2.5  flash
\multirow{12}{*}{\textbf{\texttt{gemini-2.5-flash}}}
  & $\{R_p, C_p\}$      & \multirow{2}{*}{0.9023} & 0.4975 & \multirow{2}{*}{0.8235} & 0.5022 & \multirow{2}{*}{0.8451} & 0.8636 & \multirow{12}{*}{142} & 44 & 43$^*$ \\
  & $\{R_p, C_f\}$      &                         & 0.4128 &                         & 0.4077 &                         & 0.8906 &  & 64 & 64 \\
  & $\{R_f, C_p\}$      & \multirow{2}{*}{\textbf{0.9420}} & 0.3947 & \multirow{2}{*}{\textbf{0.8887}} & 0.4201 & \multirow{2}{*}{0.9155} & 0.9474 &  & 38 & 38 \\
  & $\{R_f, C_f\}$      &                         & 0.4670 &                         & 0.4769 &                         & \textbf{0.9643} &  & 56 & 55$^*$ \\
\cmidrule(l){2-8} \cmidrule(l){10-10}
  & $E_p\{R_p, C_p\}$   & \multirow{2}{*}{0.8788} & 0.3896 & \multirow{2}{*}{0.7824} & 0.4377 & \multirow{2}{*}{0.8169} & 0.8636 &  & 44 & 44 \\
  & $E_p\{R_p, C_f\}$   &                         & 0.3677 &                         & 0.3760 &                         & 0.8281 &  & 64 & 64 \\
  & $E_p\{R_f, C_p\}$   & \multirow{2}{*}{0.9353} & 0.3364 & \multirow{2}{*}{0.8740} & 0.3577 & \multirow{2}{*}{0.9155} & 0.9474 &  & 38 & 38 \\
  & $E_p\{R_f, C_f\}$   &                         & 0.4225 &                         & 0.4320 &                         & 0.9464 &  & 56 & 56 \\
\cmidrule(l){2-8} \cmidrule(l){10-10}
  & $E_f\{R_p, C_p\}$   & \multirow{2}{*}{0.8806} & 0.3900 & \multirow{2}{*}{0.7796} & 0.4121 & \multirow{2}{*}{0.8310} & 0.9091 &  & 44 & 44 \\
  & $E_f\{R_p, C_f\}$   &                         & 0.3694 &                         & 0.3213 &                         & 0.8438 &  & 64 & 64 \\
  & $E_f\{R_f, C_p\}$   & \multirow{2}{*}{0.9371} & 0.4201 & \multirow{2}{*}{0.8733} & 0.4423 & \multirow{2}{*}{\textbf{0.9437}} & 0.9211 &  & 38 & 38 \\
  & $E_f\{R_f, C_f\}$   &                         & 0.4653 &                         & 0.4931 &                         & 0.9107 &  & 56 & 56 \\

\midrule

% % Claude-4-Haiku
\multirow{12}{*}{\textbf{\texttt{claude-4-haiku}}}
  & $\{R_p, C_p\}$      & \multirow{2}{*}{0.8485} & 0.5106 & \multirow{2}{*}{0.7255} & 0.4872 & \multirow{2}{*}{0.7887} & 0.7273 & \multirow{12}{*}{142} & 44 & 44 \\
  & $\{R_p, C_f\}$      &                         & 0.4266 &                         & 0.3619 &                         & 0.6719 &  & 64 & 63$^*$ \\
  & $\{R_f, C_p\}$      & \multirow{2}{*}{0.8550} & 0.4631 & \multirow{2}{*}{0.7413} & 0.4471 & \multirow{2}{*}{0.7887} & 0.7895 &  & 38 & 38 \\
  & $\{R_f, C_f\}$      &                         & 0.5014 &                         & 0.4813 &                         & 0.6786 &  & 56 & 56 \\
\cmidrule(l){2-8} \cmidrule(l){10-10}
  & $E_p\{R_p, C_p\}$   & \multirow{2}{*}{0.8429} & 0.5175 & \multirow{2}{*}{0.6904} & 0.5206 & \multirow{2}{*}{0.8310} & 0.7045 &  & 44 & 44 \\
  & $E_p\{R_p, C_f\}$   &                         & 0.4512 &                         & 0.4074 &                         & 0.6563 &  & 64 & 63$^*$ \\
  & $E_p\{R_f, C_p\}$   & \multirow{2}{*}{0.8592} & 0.4492 & \multirow{2}{*}{0.7183} & 0.4488 & \multirow{2}{*}{0.8592} & 0.7368 &  & 38 & 38 \\
  & $E_p\{R_f, C_f\}$   &                         & 0.4686 &                         & 0.4362 &                         & 0.6786 &  & 56 & 56 \\
\cmidrule(l){2-8} \cmidrule(l){10-10}
  & $E_f\{R_p, C_p\}$   & \multirow{2}{*}{0.8467} & 0.4548 & \multirow{2}{*}{0.7060} & 0.4313 & \multirow{2}{*}{0.8169} & 0.7727 &  & 44 & 44 \\
  & $E_f\{R_p, C_f\}$   &                         & 0.4922 &                         & 0.4658 &                         & 0.6563 &  & 64 & 63$^*$ \\
  & $E_f\{R_f, C_p\}$   & \multirow{2}{*}{0.8511} & 0.4281 & \multirow{2}{*}{0.7043} & 0.4131 & \multirow{2}{*}{0.8451} & 0.7632 &  & 38 & 38 \\
  & $E_f\{R_f, C_f\}$   &                         & 0.5053 &                         & 0.4967 &                         & 0.6786 &  & 56 & 56 \\

\bottomrule
\end{tabular}%
}

\vspace{2mm}
\parbox{\textwidth}{\centering\scriptsize Note: $E_p$ = Partial cosine enrichment, $E_f$ = Full cosine enrichment, $^*$ = Certain model results lacked empirical verifiability. $C_{(\cdot)}$ contexts are excluded for RQ2.}

\end{table*}

\subsection{LLM Detection: Interpretability and Challenges}   
We adopted seven LLMs: \texttt{gpt-4o}, \texttt{gpt-4o-mini}, \texttt{gpt-4.1-mini}, \texttt{llama-3.1-405b}, \texttt{llama-3.1-70b}, \texttt{gemini-2.5-flash}, \texttt{claude-4-haiku}, and evaluated the performance of the LLMs in classifying RQ3 based on both flaky and non-flaky observations, and RQ3 and RQ5 based solely on the ground truths of the flaky observations. The results can be found in Table~\ref{tab:results}; we employ the F1-score, Mathews Correlation Coefficient (MCC), recall, and the count of detected flaky/non-flaky tests to compare performances.

We also evaluated two additional LLMs on-site using the Ollama framework~\cite{ollama2024}. However, the results for the non-instruct tuned \texttt{llama-3.1-8b} and \texttt{llama-3.1-70b} models were excluded due to insufficient performance, as many outputs were either empty or corrupted. This underperformance is likely attributable to the fact that these are the smallest non-instruction-tuned models in our study.

Asymmetries in total score counts for combinations of the core design $\{R_p, R_f\} \times \{C_p, C_f\}$ were observed for several scenarios such as when method-level code could not be extracted, when the code was not written in Python, when no changes affected the methods, or when an artifact in one set had no corresponding entry in the other (for example, missing $R_f$ in the combinations).

\subsection{Model Performance}

\subsubsection{RQ3}

For both $R_p$ and $R_f$, \texttt{gemini-2.5-flash} ranks as best-performing with an F1-score of 0.9420 and an MCC of 0.8887. This result is somewhat counterintuitive, as \texttt{gpt-4o} is generally considered more powerful for generalized tasks. However, when only $R_p$ context is provided, \texttt{gpt-4o} outperforms all other GPT and Llama models under study with an F1 of 0.8592, nevertheless, \texttt{gpt-4.1-mini} follows-up as a more balanced result with an F1 of 0.8462 and MCC of 0.7288. This suggests that classification with limited context might be more challenging, and a more sophisticated model excels in such cases.

The \texttt{llama-3.1-405b} behaves similar to \texttt{gpt-4o}, displaying an increase in recall when given $C_f$ to $C_p$. In contrast, \texttt{llama-3.1-70b} exhibits behavior similar to \texttt{gpt-4o-mini}, with decreased performance in both contexts. We find that the more larger and newer models improve recall performances, for instance, when contrasting the RQ3 scores of 0.8592 and 0.9014 from \texttt{gpt-4o-2024-11-20} to \texttt{llama-3.1-70b} of 0.7042 and 0.6761. This suggests analyzing both natural language and code is a more complex task, requiring a more advanced model.

\subsubsection{RQ4}
Through a comparison of recall values, we assess whether code imputation at the method-level ($C_p$) or in its entirety ($C_f$) improves classification accuracy. As a general consensus, when concentrating code to the relevant problem statement, recall performance improves, which aligns with the observation that a model, similar to how a human software engineer, struggles to identify specific relevant methods to focus on when faced with the entire codebase at first glance. 

Note that we should be cautious about drawing strong conclusions when comparing $C_p$ and $C_f$, due to observation imbalance discussed in sections \ref{sec:internal_validity} and \ref{sec:external_validity}.

\subsubsection{RQ5}
Representing the most challenging task under study, requiring both multi-label classification and code analysis, we evaluate the performance scores using weighted-F1. As anticipated, the performance of all models declines compared to RQ3 and RQ4. Apart from the outlying $\{R_f, C_f\}$ RQ5 scores from \texttt{gpt-4.1-mini}, the more powerful model \texttt{gpt-4o} consistently demonstrates better performance on root cause analysis and classification.

The complexity nature of this task led to the motivation to experiment with the entablement of reasoning with \texttt{gemini-2.5-flash} which performed significantly well for majority of the research questions.

\subsection{Context Enrichment Strategies}
The original research design combinations were further extended to incorporate few-shot prompting via $E_p$/$E_f$. Results display progressive improvement in recall across majority of models for RQ3. Improvements are demonstrated in \texttt{gpt-4.0-mini} where ($R_p + E_f=+0.0468$) and ($R_f + E_f=+0.0309$). It is hypothesized that the near stagnancy and or regression of other performance metrics are due to the correlation of model task interpretability capabilities and context window saturation.

Further experiments to enrich the research design by imputing cosine ranked code snippets for RQ3 were pursued both at the method-level and entirety, however no noticeable improvements were discovered.

Based on the discussion above, we observe that $R_f$ generally aids models in making better decisions across all research questions. For RQ3 and RQ4, the performance drop is relatively small, indicating that models can still provide practical value when an issue or pull request is initially opened. Moreover, method-level code ($C_p$) appears to yield better results than full code listings ($C_f$), but further analysis is necessary, as the number of observations differs between the two setups.

\section{Threats to Validity}\label{sec:threats}

\subsection{Construct Validity}
In our study, flaky and non-flaky test labeling is a manual process that can be error-prone. To mitigate potential errors, at least two authors cross-examined all observations. We also acknowledge the limitation that our approach relies on developer-reported IRs and PRs. While this provides a reasonable basis for analysis, potential inaccuracies remain, such as misuse of terminology or misinterpretation by developers. To address this, we complemented our analysis with manual code inspections and reviews of associated fixes. However, we did not dynamically re-run the tests to observe both passing and failing executions under identical conditions, which could further validate flaky behavior. 

\subsection{Internal Validity}\label{sec:internal_validity}

The differences between methods were automatically extracted at the method level using PyDriller~\cite{Spadini2018}, which operates only on Python code. Given that Python dominates our dataset (12 out of 14 repositories are written primarily in Python), this limitation is not a significant concern. However, even in Python-focused codebases, not all repositories contain method-level data. For example, some fixes might target configuration files or global variables within Python scripts. Therefore, we capture and report the total number of observations for each experimental setup to account for such cases.

Our study performs a single forward pass for each experimental run. LLMs are inherently non-deterministic and may produce variable outputs even when inference-level hyperparameters are held constant, which can affect the consistency of observed results, especially for reasoning models. From a manual assessment of the captured outputs, we discovered that some models did not return a usable value or produced responses that could not be reliably aligned with the expected classifications. This behavior introduces additional sources of variability and limits the degree of empirical verifiability achievable under a single-pass evaluation setting.

\subsection{External Validity}\label{sec:external_validity}
Generally, software engineering studies suffer from real-world variability, and the generalization problem can only be solved partially~\cite{wieringa2015six}. One threat to external validity is the limited scope of our dataset, which, although enriched from previous studies, still focuses on a subset of quantum software repositories. As a result, the findings may not be fully representative of the broader population of quantum software projects, especially those utilizing different testing frameworks or methodologies. In future research, we hope to further expand our dataset and findings.

\section{Conclusions and Future Work}
\label{sec:conclusions}

This paper presents an automated pipeline for detecting and diagnosing quantum flaky tests by mining IRs/PRs and incorporating code context. Building on a prior dataset with 46 quantum flaky tests, which are manually detected and selected from 14 exising quantum software, we expand the dataset by identifying 25 new cases (a $\approx$54\% increase) and enriching each observation with links to associated reports, fixes, and code changes. We further evaluate a diverse set of LLMs for 1) classifying whether an IR/PR is related to a flaky test and 2) identifying the root cause of flakiness using a taxonomy grounded in prior empirical findings. The results indicate that LLMs can provide practical support for triaging flaky-test-related reports and for explaining their underlying causes, especially when prompts are constructed with appropriate textual and method-level code context.

There are several potential directions for future work. First, we plan to incorporate dynamic evidence~---~such as selective test re-execution on simulators (a typical regression testing approach adopted in quantum software development and testing)~---~to better examine suspected flaky behavior. Second, we will extend the existing pipeline to support actionable mitigation by recommending fix patterns and by exploring auto program repair techniques. Finally, we will continue expanding the dataset across additional repositories and quantum platforms, and release benchmarks that facilitate fair comparisons across detection and diagnosis approaches in quantum software engineering.

\section*{Dataset Availability}

The dataset supporting the findings of this study is publicly accessible on Zenodo at \url{https://doi.org/10.5281/zenodo.18642687} \cite{sivaloganathan_2026_18642687}.

\bibliographystyle{elsarticle-num}
\bibliography{references}

\begin{thebibliography}{10}
\expandafter\ifx\csname url\endcsname\relax
  \def\url#1{\texttt{#1}}\fi
\expandafter\ifx\csname urlprefix\endcsname\relax\def\urlprefix{URL }\fi
\expandafter\ifx\csname href\endcsname\relax
  \def\href#1#2{#2} \def\path#1{#1}\fi

\bibitem{luo2014empirical}
Q.~Luo, F.~Hariri, L.~Eloussi, D.~Marinov, An empirical analysis of flaky
  tests, in: Proceedings of the 22nd ACM SIGSOFT international symposium on
  foundations of software engineering, ACM, New York, NY, USA, 2014, pp.
  643--653.

\bibitem{micco2017state}
J.~Micco, The state of continuous integration testing@ google, in: ICST, 2017.

\bibitem{memon2017taming}
A.~Memon, Z.~Gao, B.~Nguyen, S.~Dhanda, E.~Nickell, R.~Siemborski, J.~Micco,
  Taming google-scale continuous testing, in: 2017 IEEE/ACM 39th International
  Conference on Software Engineering: Software Engineering in Practice Track
  (ICSE-SEIP), IEEE, 2017, pp. 233--242.

\bibitem{gruber2021empirical}
M.~Gruber, S.~Lukasczyk, F.~Kroi{\ss}, G.~Fraser, An empirical study of flaky
  tests in {P}ython, in: 2021 14th IEEE Conference on Software Testing,
  Verification and Validation (ICST), IEEE, IEEE, Los Alamitos, CA, USA, 2021,
  pp. 148--158.

\bibitem{lam2019idflakies}
W.~Lam, R.~Oei, A.~Shi, D.~Marinov, T.~Xie, idflakies: A framework for
  detecting and partially classifying flaky tests, in: 2019 12th ieee
  conference on software testing, validation and verification (icst), IEEE,
  2019, pp. 312--322.

\bibitem{silva2020shake}
D.~Silva, L.~Teixeira, M.~d’Amorim, Shake it! detecting flaky tests caused by
  concurrency with shaker, in: Proceedings of the 2020 IEEE International
  Conference on Software Maintenance and Evolution (ICSME), IEEE, IEEE, Los
  Alamitos, CA, USA, 2020, pp. 301--311.

\bibitem{bell2018deflaker}
J.~Bell, O.~Legunsen, M.~Hilton, L.~Eloussi, T.~Yung, D.~Marinov, {DeFlaker}:
  Automatically detecting flaky tests, in: Proceedings of the 40th
  international conference on software engineering, ACM, New York, NY, USA,
  2018, pp. 433--444.

\bibitem{alshammari2021flakeflagger}
A.~Alshammari, C.~Morris, M.~Hilton, J.~Bell, Flake{F}lagger: Predicting
  flakiness without rerunning tests, in: Proceedings of the 2021 IEEE/ACM 43rd
  International Conference on Software Engineering (ICSE), IEEE, IEEE, Los
  Alamitos, CA, USA, 2021, pp. 1572--1584.

\bibitem{verdecchia2021know}
R.~Verdecchia, E.~Cruciani, B.~Miranda, A.~Bertolino, Know you neighbor: Fast
  static prediction of test flakiness, IEEE Access 9 (2021) 76119--76134.

\bibitem{akli2023flakycat}
A.~Akli, G.~Haben, S.~Habchi, M.~Papadakis, Y.~Le~Traon, {FlakyCat}: Predicting
  flaky tests categories using few-shot learning, in: Proceedings of 2023
  IEEE/ACM International Conference on Automation of Software Test (AST), IEEE,
  IEEE, Los Alamitos, CA, USA, 2023, pp. 140--151.

\bibitem{barbosa2022test}
K.~Barbosa, R.~Ferreira, G.~Pinto, M.~d'Amorim, B.~Miranda, Test flakiness
  across programming languages, IEEE Transactions on Software Engineering
  49~(4) (2022) 2039--2052.

\bibitem{habchi2022qualitative}
S.~Habchi, G.~Haben, M.~Papadakis, M.~Cordy, Y.~Le~Traon, A qualitative study
  on the sources, impacts, and mitigation strategies of flaky tests, in:
  Proceedings of 2022 IEEE Conference on Software Testing, Verification and
  Validation (ICST), IEEE, IEEE, Los Alamitos, CA, USA, 2022, pp. 244--255.

\bibitem{zhang2023identifying}
L.~Zhang, M.~Radnejad, A.~Miranskyy, Identifying flakiness in quantum programs,
  in: 2023 ACM/IEEE International Symposium on Empirical Software Engineering
  and Measurement (ESEM), IEEE, IEEE, New York, NY, USA, 2023, pp. 1--7.

\bibitem{zhang2024automated}
L.~Zhang, A.~Miranskyy, Automated flakiness detection in quantum software bug
  reports, in: 2024 IEEE International Conference on Quantum Computing and
  Engineering (QCE), Vol.~2, IEEE, 2024, pp. 179--181.

\bibitem{kaur2025identifying}
K.~Kaur, D.~Kim, A.~Jamshidi, L.~Zhang, Identifying flaky tests in quantum
  code: A machine learning approach, arXiv preprint arXiv:2502.04471 (2025).

\bibitem{parry2021survey}
O.~Parry, G.~M. Kapfhammer, M.~Hilton, P.~McMinn, A survey of flaky tests, ACM
  Transactions on Software Engineering and Methodology (TOSEM) 31~(1) (2021)
  1--74.

\bibitem{Qiskit2019}
G.~Aleksandrowicz, et~al., Qiskit: An open-source framework for quantum
  computing, Zenodo (2019).
\newblock \href {https://doi.org/10.5281/zenodo.2562111}
  {\path{doi:10.5281/zenodo.2562111}}.

\bibitem{carleo2019netket}
G.~Carleo, K.~Choo, D.~Hofmann, J.~E. Smith, T.~Westerhout, F.~Alet, E.~J.
  Davis, S.~Efthymiou, I.~Glasser, S.-H. Lin, et~al., Netket: A machine
  learning toolkit for many-body quantum systems, SoftwareX 10 (2019) 100311.

\bibitem{sivaloganathan_2026_18642687}
J.~Sivaloganathan, A.~Jamshidi, A.~Miranskyy, L.~Zhang,
  \href{https://doi.org/10.5281/zenodo.18642687}{Automating detection and
  root-cause analysis of flaky tests in quantum software — supplementary
  dataset} (Feb. 2026).
\newblock \href {https://doi.org/10.5281/zenodo.18642687}
  {\path{doi:10.5281/zenodo.18642687}}.
\newline\urlprefix\url{https://doi.org/10.5281/zenodo.18642687}

\bibitem{fatima2022flakify}
S.~Fatima, T.~A. Ghaleb, L.~Briand, Flakify: A black-box, language model-based
  predictor for flaky tests, IEEE Transactions on Software Engineering 49~(4)
  (2022) 1912--1927.

\bibitem{muennighoff2022mteb}
N.~Muennighoff, N.~Tazi, L.~Magne, N.~Reimers,
  \href{https://arxiv.org/abs/2210.07316}{Mteb: Massive text embedding
  benchmark}, arXiv preprint arXiv:2210.07316 (2022).
\newblock \href {https://doi.org/10.48550/ARXIV.2210.07316}
  {\path{doi:10.48550/ARXIV.2210.07316}}.
\newline\urlprefix\url{https://arxiv.org/abs/2210.07316}

\bibitem{emb2024mxbai}
S.~Lee, A.~Shakir, D.~Koenig, J.~Lipp,
  \href{https://www.mixedbread.ai/blog/mxbai-embed-large-v1}{Open source
  strikes bread - new fluffy embeddings model} (2024).
\newline\urlprefix\url{https://www.mixedbread.ai/blog/mxbai-embed-large-v1}

\bibitem{li2023angle}
X.~Li, J.~Li, Angle-optimized text embeddings, arXiv preprint arXiv:2309.12871
  (2023).

\bibitem{wolf2019huggingface}
T.~Wolf, Huggingface's transformers: State-of-the-art natural language
  processing, arXiv preprint arXiv:1910.03771 (2019).

\bibitem{SFRAIResearch2024}
S.~R.~J. Rui~Meng, Ye~Liu,
  \href{https://blog.salesforceairesearch.com/sfr-embedded-mistral/}{Sfr-embedding-mistral:enhance
  text retrieval with transfer learning}, Salesforce AI Research Blog (2024).
\newline\urlprefix\url{https://blog.salesforceairesearch.com/sfr-embedded-mistral/}

\bibitem{wang2024improving}
L.~Wang, N.~Yang, X.~Huang, L.~Yang, R.~Majumder, F.~Wei, Improving text
  embeddings with large language models, in: Proceedings of the 62nd Annual
  Meeting of the Association for Computational Linguistics (Volume 1: Long
  Papers), 2024, pp. 11897--11916.

\bibitem{wang2022text}
L.~Wang, N.~Yang, X.~Huang, B.~Jiao, L.~Yang, D.~Jiang, R.~Majumder, F.~Wei,
  Text embeddings by weakly-supervised contrastive pre-training, arXiv preprint
  arXiv:2212.03533 (2022).

\bibitem{macqueen1967some}
J.~Macqueen, Some methods for classification and analysis of multivariate
  observations, in: Proceedings of 5-th Berkeley Symposium on Mathematical
  Statistics and Probability/University of California Press, University of
  California Press, Berkeley, CA, 1967.

\bibitem{dutta2020detecting}
S.~Dutta, A.~Shi, R.~Choudhary, Z.~Zhang, A.~Jain, S.~Misailovic, Detecting
  flaky tests in probabilistic and machine learning applications, in:
  Proceedings of the 29th ACM SIGSOFT international symposium on software
  testing and analysis, 2020, pp. 211--224.

\bibitem{trivsovic2025rapid}
A.~Tri{\v{s}}ovi{\'c}, A.~Fogelson, J.~Sivaloganathan, N.~Thompson, The rapid
  growth of ai foundation model usage in science, arXiv preprint
  arXiv:2511.21739 (2025).

\bibitem{sivaloganathan2025mapping}
J.~Sivaloganathan, A.~Tri{\v{s}}ovi{\'c}, N.~Thompson, Mapping the impact of
  foundation models on the un sustainable development goals, in: 2025 IEEE
  International Conference on eScience (eScience), IEEE, 2025, pp. 375--376.

\bibitem{fatima2024flakyfix}
S.~Fatima, H.~Hemmati, L.~Briand, Flakyfix: Using large language models for
  predicting flaky test fix categories and test code repair, IEEE Transactions
  on Software Engineering (2024).

\bibitem{LangChai97:online}
H.~Chase, \href{https://github.com/langchain-ai/langchain}{{LangChain}} (Oct.
  2022).
\newline\urlprefix\url{https://github.com/langchain-ai/langchain}

\bibitem{dubey2024LLaMA}
A.~Dubey, A.~Jauhri, A.~Pandey, A.~Kadian, A.~Al-Dahle, A.~Letman, A.~Mathur,
  A.~Schelten, A.~Yang, A.~Fan, et~al., The llama 3 herd of models, arXiv
  e-prints (2024) arXiv--2407.

\bibitem{openai2024gpt4omini}
I.~OpenAI, Gpt-4o mini: Advancing cost-efficient intelligence,
  \url{https://openai.com/index/gpt-4o-mini-advancing-cost-efficient-intelligence/}
  (2024).

\bibitem{openai2024gpt4o}
I.~OpenAI, Hello gpt-4o, \url{https://openai.com/index/hello-gpt-4o/} (2024).

\bibitem{openai2025gpt41}
{OpenAI, Inc.}, Introducing gpt-4.1 in the api,
  \url{https://openai.com/index/gpt-4-1/} (2025).

\bibitem{google2025gemini25flash}
{Google DeepMind / Google AI}, Gemini 2.5 flash: A cost-efficient multimodal
  reasoning model, \url{https://ai.google.dev/gemini-api/docs/models/gemini}
  (2025).

\bibitem{anthropic2025claudehaiku}
{Anthropic, Inc.}, Claude haiku 4.5: A fast, cost-efficient model in the claude
  family, \url{https://www.anthropic.com/claude/haiku} (2025).

\bibitem{Overview51:online}
{OpenAI}, \href{https://platform.openai.com/docs/overview}{Overview - openai
  api} (2024).
\newline\urlprefix\url{https://platform.openai.com/docs/overview}

\bibitem{AnthropicAPIOverview:online}
{Anthropic}, \href{https://docs.anthropic.com/en/api/overview}{Api overview –
  anthropic documentation} (2025).
\newline\urlprefix\url{https://docs.anthropic.com/en/api/overview}

\bibitem{VertexAI96:online}
{Google Cloud}, \href{https://cloud.google.com/vertex-ai/docs}{Vertex ai
  documentation | google cloud} (2024).
\newline\urlprefix\url{https://cloud.google.com/vertex-ai/docs}

\bibitem{feng2020codebert}
Z.~Feng, D.~Guo, D.~Tang, N.~Duan, X.~Feng, M.~Gong, L.~Shou, B.~Qin, T.~Liu,
  D.~Jiang, et~al., Codebert: A pre-trained model for programming and natural
  languages, arXiv preprint arXiv:2002.08155 (2020).

\bibitem{ollama2024}
J.~Morgan, M.~Chiang, \href{https://github.com/ollama/ollama}{{Ollama}} (2024).
\newline\urlprefix\url{https://github.com/ollama/ollama}

\bibitem{Spadini2018}
D.~Spadini, M.~Aniche, A.~Bacchelli, {PyDriller: Python framework for mining
  software repositories}, in: Proceedings of the 2018 26th ACM Joint Meeting on
  European Software Engineering Conference and Symposium on the Foundations of
  Software Engineering - ESEC/FSE 2018, ACM, New York, New York, USA, 2018, pp.
  908--911.
\newblock \href {https://doi.org/10.1145/3236024.3264598}
  {\path{doi:10.1145/3236024.3264598}}.

\bibitem{wieringa2015six}
R.~J. Wieringa, M.~Daneva, Six strategies for generalizing software engineering
  theories, Science of computer programming 101 (2015) 136--152.
\newblock \href {https://doi.org/10.1016/j.scico.2014.11.013}
  {\path{doi:10.1016/j.scico.2014.11.013}}.

\end{thebibliography}

\end{document}